\documentclass[aps,twocolumn,showpacs,nofootinbib]{revtex4-1}
\usepackage{graphicx}   
\usepackage{latexsym}   
\usepackage{enumerate}
\usepackage{rotating,booktabs,multirow}
\usepackage{amssymb}
\usepackage{color}
\usepackage{hyperref}

\begin{document}
\title{Kinetic mass shifts of $\rho$(770) and $K^*$(892) in Au+Au reactions at $E_\mathrm{beam}=1.23~A$GeV}
\author{Tom~Reichert$^{1,3}$, Marcus~Bleicher$^{1,2,3}$}

\affiliation{$^1$ Institut f\"ur Theoretische Physik, Goethe Universit\"at Frankfurt, Max-von-Laue-Strasse 1, D-60438 Frankfurt am Main, Germany}
\affiliation{$^2$ GSI Helmholtzzentrum f\"ur Schwerionenforschung GmbH, Planckstr. 1, 64291 Darmstadt, Germany}
\affiliation{$^3$ Helmholtz Research Academy Hesse for FAIR (HFHF), Campus Frankfurt, Max-von-Laue-Str. 12, 60438 Frankfurt, Germany}

\begin{abstract}
We explore the properties of the last generation of $\rho$(770) and $K^*$(892) resonances created in Au+Au reactions at E$_\mathrm{lab}=1.23$~$A$GeV via the reconstruction in the hadronic channel using the UrQMD model. Such resonance studies allow to obtain detailed information on the emission source in such reactions in a complementary way to di-lepton or HBT studies. We observe a freeze-out hierarchy and predict mass shifts for the $\rho$(770) resonance ($\Delta m_\rho \approx -330$ MeV) and the $K^*$(892) resonance ($\Delta m_{K^*} \approx -30$ MeV). These mass shifts are related to the regeneration cycles of each resonance species and are of purely kinetic origin. Experimentally our predictions can be tested using the GSI-HADES experiment or the lowest RHIC-BES energy.
\end{abstract}

\maketitle
\section{Introduction}
The study of subatomic matter is one of the main research topics of modern physics. The most prominent force acting on this length scale is the strong interaction which is described by Quantum Chromo Dynamics (QCD). On earth, QCD can be best explored at particle accelerators, using the collision of heavy ions. At moderate collision energies, as discussed here, the created matter reaches densities three to four times higher than normal nuclear density (often dubbed the ''high density frontier'') and temperatures in the range of neutron star mergers. The other extreme (high temperature frontier) is explored in today's biggest accelerators, like the LHC or RHIC, to study the deconfined phase consisting of free quarks and gluons. A third domain exploring the onset of deconfinement will be extensively studied in future facilities like NICA or CBM at FAIR. There one aims for baryon density-temperature combinations to reach the mixed phase and to explore the QCD phase transition, the critical point and further exotic phases like Quarkyonic matter. This goes hand in hand with modifications of the spectral functions of resonances, their widths or masses. The investigation of mass shifts or broadening of the spectral function of hadrons has already a long history \cite{Pisarski:1981mq}. Most prominent has been the discussion of the mass shift of the $\rho$ meson (Brown-Rho-Scaling \cite{Brown:1991kk}) that was finally resolved by the NA60 experiment in In+In reactions at E$_\mathrm{lab}=160$~AGeV \cite{Damjanovic:2006bd,Damjanovic:2007qm,Arnaldi:2006jq}. As has been clearly shown in Refs. \cite{Rapp:2010sj,Rapp:2004zh}, at this energy chiral symmetry and in-medium interactions do mainly lead to a broadening of the $\rho$-spectral function, as observed in the di-lepton channel, without a mass shift. 

While di-leptons are excellent probes for vector meson properties they cannot be used to explore the mass distribution of other hadrons that do not carry photon quantum numbers. Here, n-particle hadron correlations need to be studied to reconstruct the invariant mass spectrum of the resonance under exploration. Extensive pioneering studies in nucleus-nucleus reactions were done at the CERN-SPS (Pb+Pb, E$_\mathrm{lab}=160$~AGeV) \cite{Friese:2002re,Markert:2002rw} and later at RHIC (Au+Au, $\sqrt{s_\mathrm{NN}}=200$~GeV) \cite{Markert:2004xx,Markert:2005ms} and continued at LHC (Pb+Pb, $\sqrt{s_\mathrm{NN}}=2.76$~TeV and 7~TeV) \cite{Badala:2007zza,Khuntia:2020aan}. Together with a systematic theoretical investigation \cite{Torrieri:2001ue,Rafelski:2001hp,Bleicher:2002dm,Shuryak:2002kd,Pratt:2003vb,Vogel:2005pd,Badala:2012hjj,Aichelin:2015nqc,Knospe:2015nva,Knospe:2021jgt}, these studies led to a deeper understanding of the complex absorption and regeneration interplay behind the suppression patterns of various hadron resonances. Unfortunately, at lower energies similar studies are experimentally more difficult to accomplish. The reason is that the mixed event method needs substantial statistics and broad phase space coverage.

On the experimental side, pioneering studies in the 1-2~AGeV energy regime were performed by the FOPI collaboration \cite{Hong:1997ka,Eskef:2001qg}. Already FOPI suggested a mass shift of the $\Delta$ resonances in nucleus-nucleus collisions, if reconstructed in the hadronic channel. This has recently been confirmed by the HADES experiment  \cite{Adamczewski-Musch:2020edy} and interpreted as a regeneration effect near the kinetic freeze-out surface \cite{Reichert:2019lny,Reichert:2019zab}. The recent HADES data on $\Delta^{++}$ production has demonstrated the excellent capacities of the HADES experiment for 2-particle correlation measurements. With this now established experimental tool, we want to propose further pioneering studies into other hadron resonances at low energies. These studies will then allow to bridge to the data on hadronically reconstructed $\rho$(770) and $K^{*}$(892) meson resonances which have been under detailed investigation at RHIC (BES) and LHC energies for more than a decade.

Here we propose to extend previous studies on the connection between the kinetic decoupling stage and the invariant mass of hadronically reconstructed resonances. Two meson resonances are prime candidates for such a study: $\rho$ mesons and $K^*$ resonances. In the pion channel the $\rho$(770), decaying into $\pi^+\pi^-$ is a most interesting hadron resonance, because it has already been successfully measured in its di-lepton channel at SIS energies \cite{Adamczewski-Musch:2019byl} and numerous theoretical studies of its decay into di-leptons have been performed \cite{Bratkovskaya:2007jk,Endres:2015fna,Larionov:2020fnu}. As a short lived resonance, the $\rho$ decays shortly after it was formed and undergoes multiple regeneration cycles until freeze-out. Further measurements in the hadronic channel allow complement the di-lepton studies with independent observations and to map out the decoupling stage in more detail. In addition, direct measurements of the $\rho$ are important to benchmark simulation models.

For the exploration of strange resonances the $K^*$(892) meson is a well explored hadron resonance at higher energies. It is also very interesting at low energies, because it can still be abundantly produced. From the physics point of view, its investigation is promising because it was shown that the spectral function of the $K^*$ resonances may provide information about the early stage of a collision \cite{Ilner:2016xqr,Ilner:2017tab}.  In contrast to higher energies the regeneration of $K^*$ is strongly suppressed at lower collision energies. Due to the low temperatures, the typical invariant mass of a thermal pion and a thermal kaon is off the mass peak of the $K^*$ spectral function preventing regeneration (regeneration in the baryon channel is not possible due to the lack of a resonance in the $K+p$ channel) \cite{Vogel:2006rm}. This results in an earlier kinetic freeze-out of $K^*$'s than e.g. $\Delta(1232)$'s or the $\rho$'s. A comparative analysis of the hadronically reconstructed $\rho$(770) and $K^*$(892) invariant mass distributions could therefore provide additional information on the dynamics and properties of the freeze-out stage and on the relative regeneration strength of these resonances.

\subsection{The UrQMD model and the reconstruction of resonances}
\subsubsection{The UrQMD model}
For the purpose of this investigation, the Ultra-relativistic Quantum Molecular Dynamics (UrQMD) transport model \cite{Bass:1998ca,Bleicher:1999xi,Bleicher:2022kcu} is used.  UrQMD is based on the covariant propagation of hadrons and their interactions by elastic and/or inelastic collisions. The interaction cross sections are extracted from experimental data or derived using effective models. Within the UrQMD hadronic transport approach mass dependent decay widths are employed. UrQMD includes mesonic and baryonic resonances which bridge to string excitations and decays at higher energies. While UrQMD includes potential interactions to allow simulations of different equations-of-state, for the present study, we employ the cascade mode (no potential interactions) due to the large amount of statistics needed. Collisional broadening, this means a modification of the total width of the hadron due to interactions with the medium, is always included in the simulations. Especially for di-lepton production collisional broadening is important to obtain correct di-lepton rates at high hadron densities. In UrQMD, collisional broadening has been studied extensively with respect to di-lepton production, see e.g. \cite{Schumacher:2006wc,Vogel:2007yu}, however, it was also shown in \cite{Vogel:2007yu}, that collisional broadening is not important at the low densities relevant for the present investigation. The model is well tested in the HADES energy regime, for model results at these energies, we refer the reader to  \cite{Sombun:2018yqh,Hillmann:2019wlt,Reichert:2020uxs}.

\subsubsection{Resonance reconstruction\label{reco}}
During the time evolution of the collision mesonic and baryonic resonances arise and decay again until kinetic freeze-out where even the elastic scattering rates cease and the system breaks up \cite{Reichert:2020oes}. Depending on the life time of the resonance, one can roughly separate the short lived and long lived resonances: If the resonance is long lived it can emerge from the hot and dense region because it will not decay until kinetic freeze-out (this is typically the case for the $\phi$ meson). If the resonance is short lived it is mainly produced near to the kinetic freeze-out and represents therefore only the last generation of a series of short lived resonances going through decay and regeneration cycles (e.g. the $\Delta$ resonance). Resonances from the late stages are usually formed by s-channel processes between produced hadrons e.g. $\pi^++\pi^-\rightarrow\rho^0$ at high energies, while at low energies discussed here the decay of excited baryon resonance states dominates light meson production. In order to measure a baryon or meson resonance, the resonance has to be reconstructed in the experiment. This is experimentally done by an analysis of the invariant mass distribution in the resonances decay channels, e.g. the $\Delta^{++}$ has been observed via the $\pi^++p$ channel \cite{Adamczewski-Musch:2020edy} by the HADES collaboration. To obtain the signal one subtracts the uncorrelated pairs from the correlated pairs in a mixed event analysis. This method requires a substantial amount of statistics and is error-prone (different fits to the background shape, acceptance problems that may result in artificial structures, different efficiencies, etc.) \cite{Drijard:1984pe}. 
\begin{figure} [t!]
	\includegraphics[width=\columnwidth]{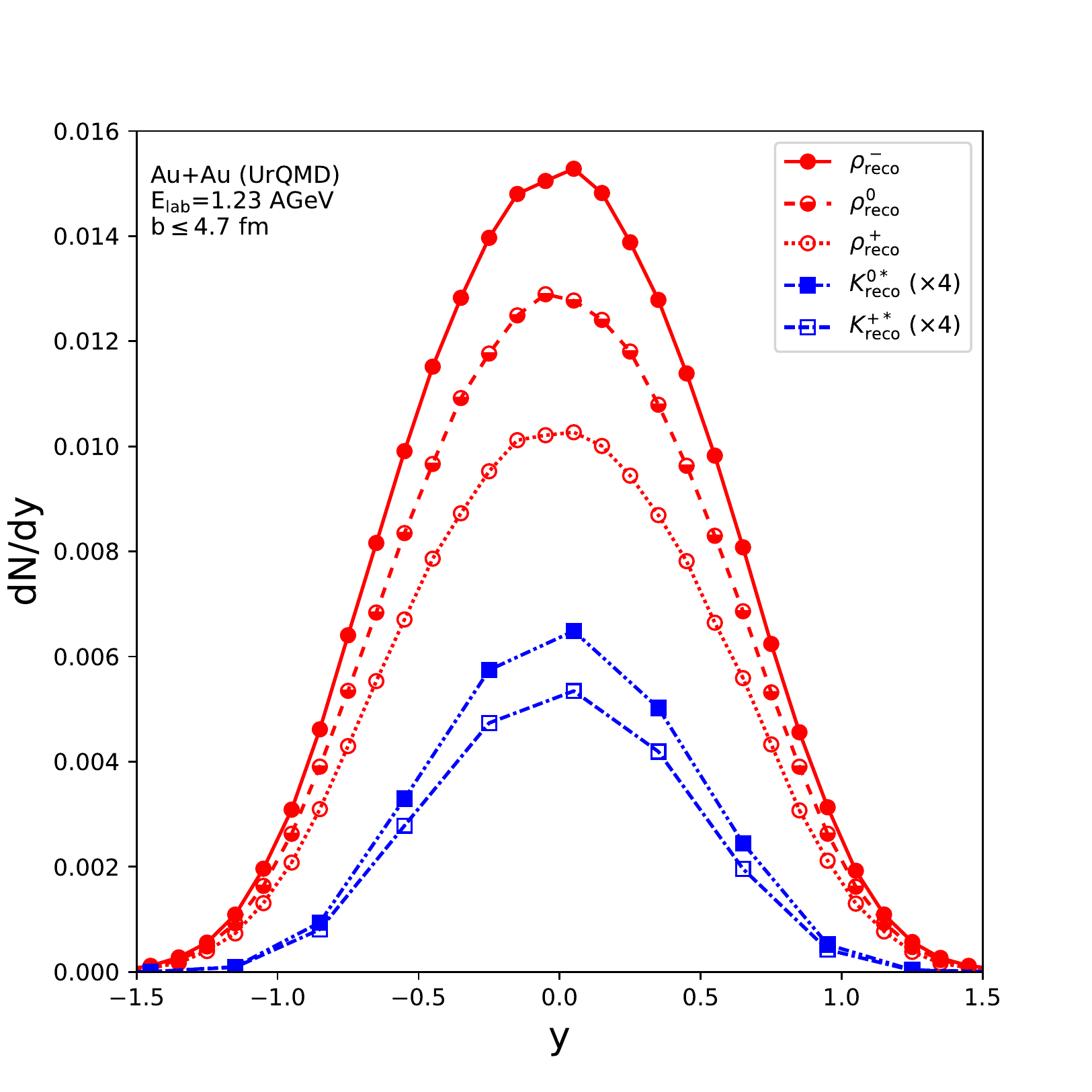}
	\caption{[Color online] Rapidity distributions of reconstructable $\rho^-$ (solid red line, full circles), $\rho^0$ (dashed red line, half-filled circles), and  $\rho^+$ (red dotted line, open circles) and $K^{0*}$ (blue dot-dot-dashed line, full squares), $K^{+*}$ (blue dash-dash-dotted line, open squares) from central Au+Au reactions at $E_\mathrm{beam}=1.23~A$GeV from UrQMD.}\label{dndy_iso}
\end{figure}
\begin{table} [b!]
	\centering
	\begin{tabular}{|r|l|c|c|}
		\hline
		Resonance & Decay channel & BR & Invariant mass range \\ \hline 
		\hline
		$\rho^0$(770) & $\rho^0\rightarrow\pi^++\pi^-$ & 100\% & 0.270 -- 1.0~GeV \\
		$K^{0*}$(892) & $K^{0*}\rightarrow K^+ + \pi^-$ & 100\% & 0.629 -- 1.1~GeV \\
		\hline
	\end{tabular}
	\caption{Possible 2-particle decay channels of the neutral $\rho$(770) and $K^*$(892) resonances, their individual branching ratios and the invariant mass range used to determine the hadron yields.}\label{table:decay-channels}
\end{table}

On the theory side the same method can of course be applied. However, one usually uses a different approach as described in \cite{Bleicher:2002dm,Reichert:2019lny,Reichert:2019zab,Steinheimer:2015sha}. In this method, every decaying resonance in the simulation is identified and each of its daughter particles is further tracked through the calculation until it reacts again or reaches the kinetic freeze-out surface. This allows to calculate for each individual decaying resonance, if it can be reconstructed and observed experimentally. The following reconstruction scenarios that result in an observable resonance are considered: 
\begin{enumerate}
	\item 	Both daughter particles do not interact further until they reach kinetic freeze-out. 
	\item  	One daughter particle rescatters elastically until it reaches kinetic freeze-out, the other daughter particle escapes the system with unchanged momentum. 
	\item 	Both daughter particles rescatter elastically until kinetic freeze-out.
\end{enumerate}
All other cases, i.e. daughter particles with inelastic reactions, do not allow to reconstruct the resonances due to signal loss in the invariant mass distribution \cite{Reichert:2019lny}.

In this paper, we explore the $\rho$(770) and $K^*$(892) resonances to provide complementary information to the already measured and theoretically analyzed $\Delta$ resonance. In the case of the $\rho$(770) meson we focus on the neutral $\rho^{0}$, because both of its decay daughters carry electric charge and are thus well measurable by the HADES experiment. In the case of the $K^{*}$(892) also the neutral $K^{0*}$(892) is of main interest due to its charged decay channel. From here on, we will drop the round brackets with the mass number of each resonance for the sake of brevity. To obtain the specific particle yields the invariant mass spectra are summed up over a specific mass region as shown in Table \ref{table:decay-channels}.

\section{Results}
\subsection{Rapidity and transverse momentum spectra}
Let us start with the rapidity distributions of the $K^*$ and $\rho$ meson resonances. Fig. \ref{dndy_iso} shows the rapidity distributions of reconstructable $\rho^-$ (solid red line, full circles), $\rho^0$ (dashed red line, half-filled circles), and  $\rho^+$ (red dotted line, open circles) and $K^{0*}$ (blue dot-dot-dashed line, full squares), $K^{+*}$ (blue dash-dash-dotted line, open squares). One observes that also in the resonances, the initial isospin unbalance of the gold nuclei is reflected in the enhancement of more negatively charged isospin states. 
\begin{figure} [t!]
	\includegraphics[width=\columnwidth]{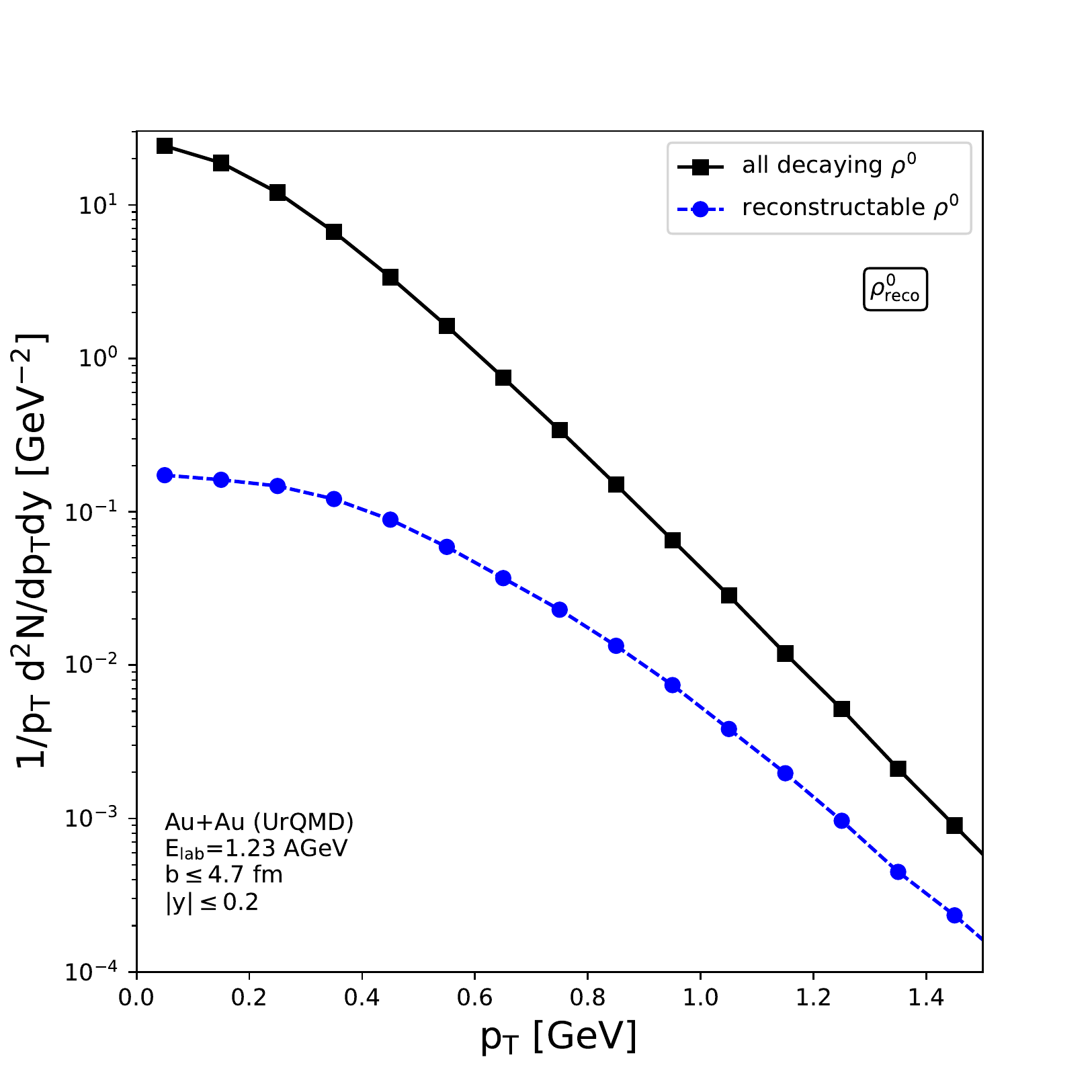}
	\caption{Transverse momentum distributions at mid-rapidity of reconstructed $\rho^0$ resonances (blue line with full circles) and of all decaying resonances (black line with full squares) in Au+Au reaction at $E_\mathrm{beam}=1.23~A$GeV from UrQMD.}\label{rho_dndpt}
\end{figure}
\begin{figure} [t!]
	\includegraphics[width=\columnwidth]{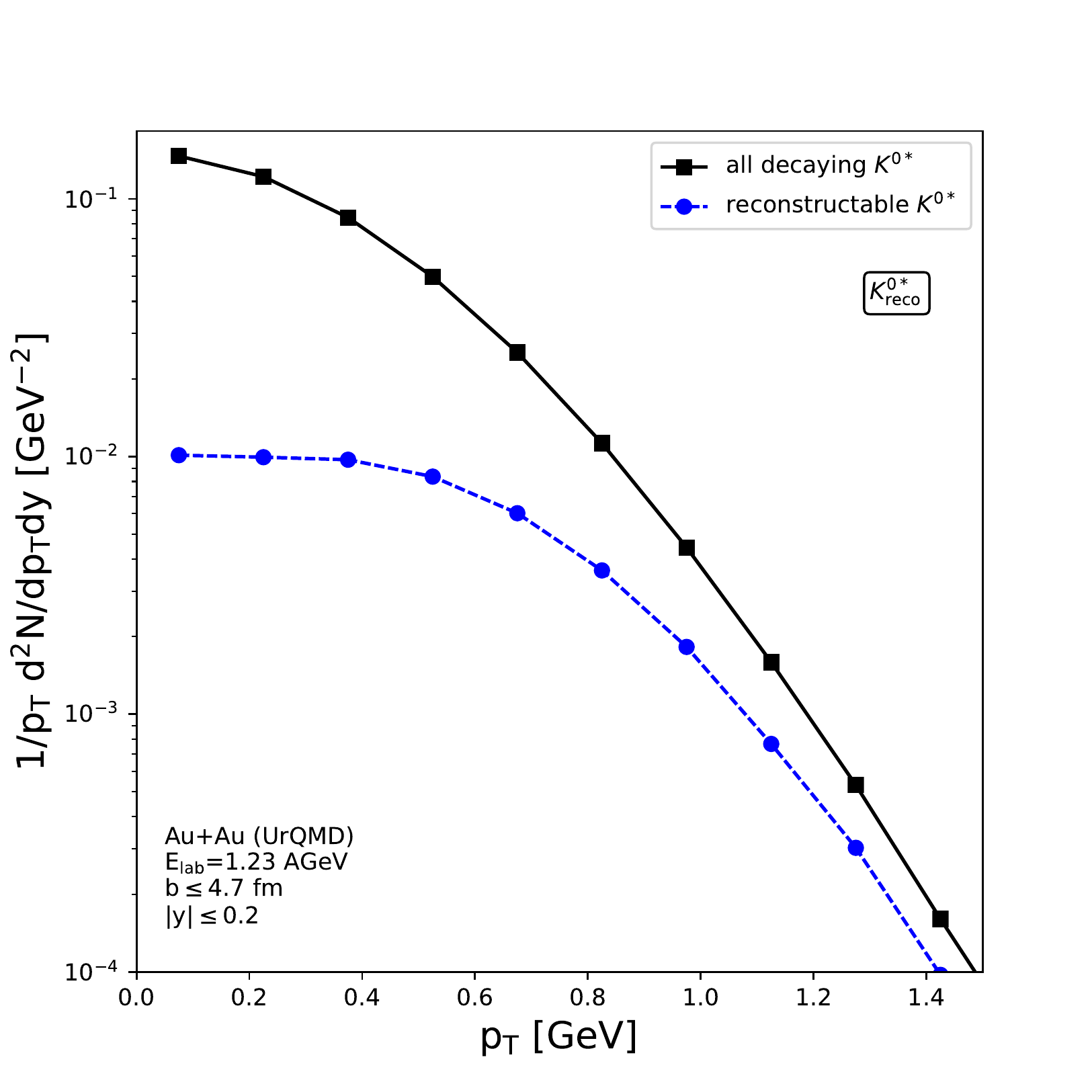}
	\caption{Transverse momentum distributions at mid-rapidity of reconstructed $K^{0*}$ resonances (blue line with full circles) and of all decaying resonances (black line with full squares) in Au+Au reaction at $E_\mathrm{beam}=1.23~A$GeV from UrQMD.}\label{kstar_dndpt}
\end{figure}

Next, we explore the transverse momentum spectra of the resonances. Figs. \ref{rho_dndpt} and \ref{kstar_dndpt} show the transverse momentum spectra for reconstructable $\rho^0$ and $K^{0*}$ resonances, respectively at mid-rapidity (dashed line with full blue circles) and all decaying resonances (dashed line with full black squares). One observes that in comparison to all decaying resonances, the transverse momentum spectrum of resonances  reconstructed in the final state hadronic channel is suppressed at low $p_{\rm T}$. The greatest suppression can be observed in the reconstructed $\rho$ meson at zero transverse momentum where the yield is suppressed by up to a factor of $\approx$~100. The yield is rather slowly approaching the yield of all decaying $\rho$ resonances. Similarly, the reconstructed $K^*$ show a suppression by up to a factor of $\approx$~10 towards vanishing $p_{\rm T}$. This observation clearly indicates that at low transverse momenta both resonances decay inside the bulk of the system which leads to a strong rescattering of the slow daughter particles. This intense interaction with the surrounding medium establishes a freeze-out hierarchy \cite{Reichert:2022qvt}: Resonances with larger interaction cross sections of their daughter particles in baryon rich matter are shifted to the end of the kinetic freeze-out, while resonances with smaller interaction cross sections of their daughter particles in baryon rich matter are reconstructed from an earlier reaction time. Thus reconstructable $K^*$ should emerge from a hotter region of the fireball than hadronically reconstructable $\rho$ mesons.

Turning to higher transverse momenta we observe that the fraction of resonances that can be reconstructed increases, suggesting that these resonances on average decouple earlier than at low transverse momenta (see also \cite{Vogel:2009kg} for a detailed study). In the case of the $\rho$ meson, the high $p_\mathrm{T}$ resonances are still suppressed roughly by a factor of 10 indicating that they still participate to some extent with the nuclear medium. The $K^*$ however shows only a minor suppression by roughly a factor of 2 at large transverse momenta. This finding states that many $K^*$ formed at high $p_\mathrm{T}$ (or very early) do not undergo significant rescattering patterns. 

\subsection{Invariant mass distribution}
How can we test this freeze-out hierarchy? Following \cite{Reichert:2020uxs} we explore the freeze-out hierarchy in $p_\mathrm{T}$ using the invariant mass distribution of the resonances. Using the dynamical UrQMD simulations, which include resonance production, decays and regeneration allows to get the full space-time differential emission pattern to reconstruct the resonances. For the $\rho$ resonances one finds that the final emission is shifted (as expected) to a rather late stage in the evolution for the following reason: Due to the large cross section of the daughter particles and its very short life time, the $\rho$ is captured in a regeneration cycle like $\pi^+\pi^-\leftrightarrow\rho^{0}$ or via the decay of an excited baryon state, e.g. $N^*\leftrightarrow N+\rho$. The relative importance of the two direct $\rho$-cycles depend on the baryon density via $\mu_{\rm B}$ and the temperature of the system. At the energy under consideration here, with high $\mu_{\rm B}$ and a rather low temperature $T$, the baryonic cycle dominates. It is therefore clear that $\rho$ mesons will be recreated throughout the whole evolution of the system and can only decouple when the expansion rate exceeds the large scattering rate \cite{Inghirami:2021zja}, i.e. the kinetic freeze-out hyper-surface is reached. 

In contrast, strange meson resonances like the $K^*$ can provide information about an earlier stage of the system. With a width of only 47~MeV \cite{Tanabashi:2018oca} the $K^*$ will decay after $\approx$~4~fm/c into a $\pi+K$. Due to its small cross section in baryon rich matter, the daughter kaon will leave the system mainly undisturbed. The $K^*$ regeneration cycle is therefore suppressed at low temperatures and high baryon densities, like in the system under investigation here. The effect of the missing $K^*$ regeneration cycle will allow to look into earlier stages of the reaction with $K^*$'s as compared to $\rho$ mesons.

The different regimes of emission of reconstructable resonances leads to different populations of the spectral function of the hadron resonance reflecting the local temperatures at the emission point of the resonance (i.e. the space-time point at which the parent resonance decays into its daughter particles). As discussed extensively in \cite{Reichert:2019lny,Reichert:2020uxs}, low temperatures result in a population of the low mass tail of the spectral function, while very high temperatures might even lead to an overpopulation of masses above the peak mass. Thus, the invariant mass distribution of the reconstructed $\rho^{0}$ and $K^{0*}$ mesons should to first approximation mirror the temperature of the system at the kinetic decoupling surface of each of these hadron resonances. The mass of a resonance going through many decay/regeneration cycles (like the $\rho$) should hence be drastically reduced due to the very low emission temperatures, while the mass of a $K^*$ should receive only a minor shift, because the $K^*$ is emitted from an earlier and hotter time of the reaction. For both cases, we also expect that higher transverse momenta show less mass shift, because these daughter particles decouple on average earlier from the system. Following \cite{Reichert:2019lny,Reichert:2019zab}, we approxiamte the invariant mass distributions by a Breit-Wigner (BW) spectral function multiplied by the thermal weight of each resonance (often called phase-space factor (PS)) \cite{Ilner:2016xqr,Ilner:2017tab}. I.e. 
\begin{equation}\label{BWxPS}
P(m_i,p_{\mathrm{T}}) \propto BW(m_i,m_i^0,\Gamma_{\rm tot}(m_i)) \times PS(m_i,p_{\mathrm{T}},T)
\end{equation}
with 
\begin{equation}\label{BW}
BW(m_i,m_i^0,\Gamma_i^0)\propto\frac{\Gamma_{\rm tot}(m_i) m_{i}}{\left(m_{i}^2-m_{i}^0\right)^2+(\Gamma_{\rm tot}(m_i) m_{i})^2}
\end{equation} 
with $\Gamma_{\rm tot}(m_i)=\sum \Gamma_{i\rightarrow j,k}(m_i)$ using 
\begin{equation}\label{eq:width}
    \Gamma_{i\rightarrow j,k}(m_i) = \Gamma_i^0 \frac{m_i^0}{m_i} \left(\frac{\langle p_{j,k}(m_i)\rangle}{\langle p_{j,k}(m_i^0)\rangle}\right)^{2l+1} \frac{1.2}{1 + 0.2\left(\frac{\langle p_{j,k}(m_i)\rangle}{\langle p_{j,k}(m_i^0)\rangle}\right)^{2l}}
\end{equation}
with $p_{j,k}$ being the momenta of the decay products in the resonance rest frame and $l$ being the resonance's angular momentum and 
\begin{equation}\label{PS}
PS(m_i,p_{\mathrm{T}},T)\propto\frac{ m_{i}}{\sqrt{m_i^2+p_{\mathrm{T}}^2}}\exp\left(-\frac{\sqrt{m_i^2+p_{\mathrm{T}}^2}}{T}\right).
\end{equation}
In Eqs. (\ref{BWxPS})-(\ref{PS}) $m_i$ is the invariant mass of resonance $i$, $m_i^0$ describes the nominal mass of resonance $i$, $\Gamma_{i\rightarrow j,k}(m_i)$ is the partial decay width (in our case identical to the total decay width) of resonance $i$ decaying into particles $j$ and $k$ which depends on the vacuum decay width $\Gamma_i^0$ whose value is taken from the Particle Data Group \cite{Tanabashi:2018oca}, $p_{\rm T}$ is the transverse momentum and $T$ denotes the local emission temperature. In line with our discussion, we observe two features from the qualitative BW$\times$PS formula: a) the average mass of a resonance increases with increasing transverse momentum and b) the average mass of a resonance decreases with decreasing temperature. 

\begin{figure} [t!]
	\includegraphics[width=\columnwidth]{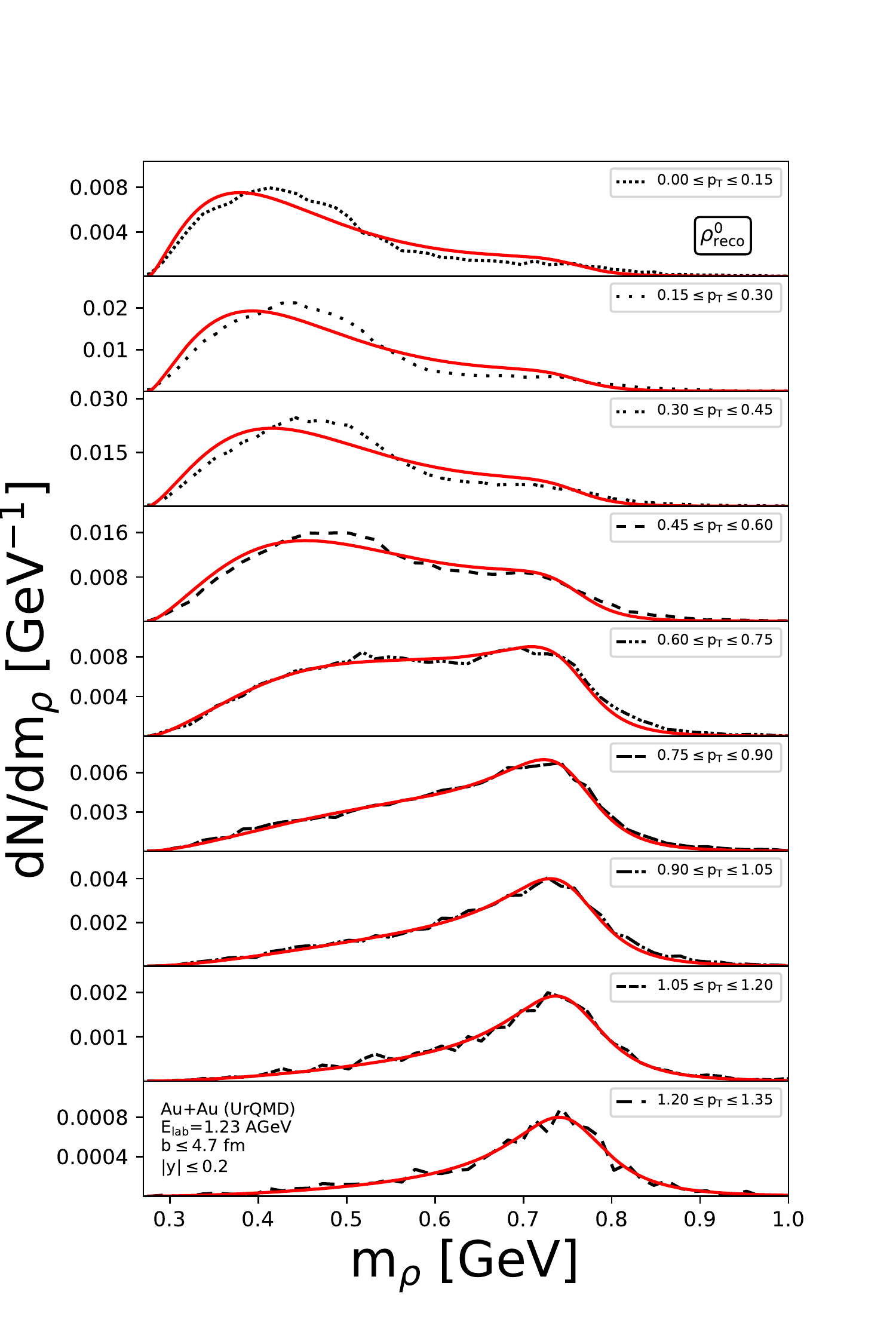}
	\caption{Invariant mass distribution of the reconstructable $\rho^0$ mesons (black lines) for several transverse momentum bins ranging from 0 to 0.15~GeV at the top to 1.2 to 1.35~GeV at the bottom at mid-rapidity in central Au+Au collisions at $E_\mathrm{beam}=1.23~A$GeV from UrQMD. Also shown are fits to the distributions following the BW$\times$PS argument as red lines.\label{rho_dndm}}
\end{figure}

\begin{figure} [t!hb]
	\includegraphics[width=\columnwidth]{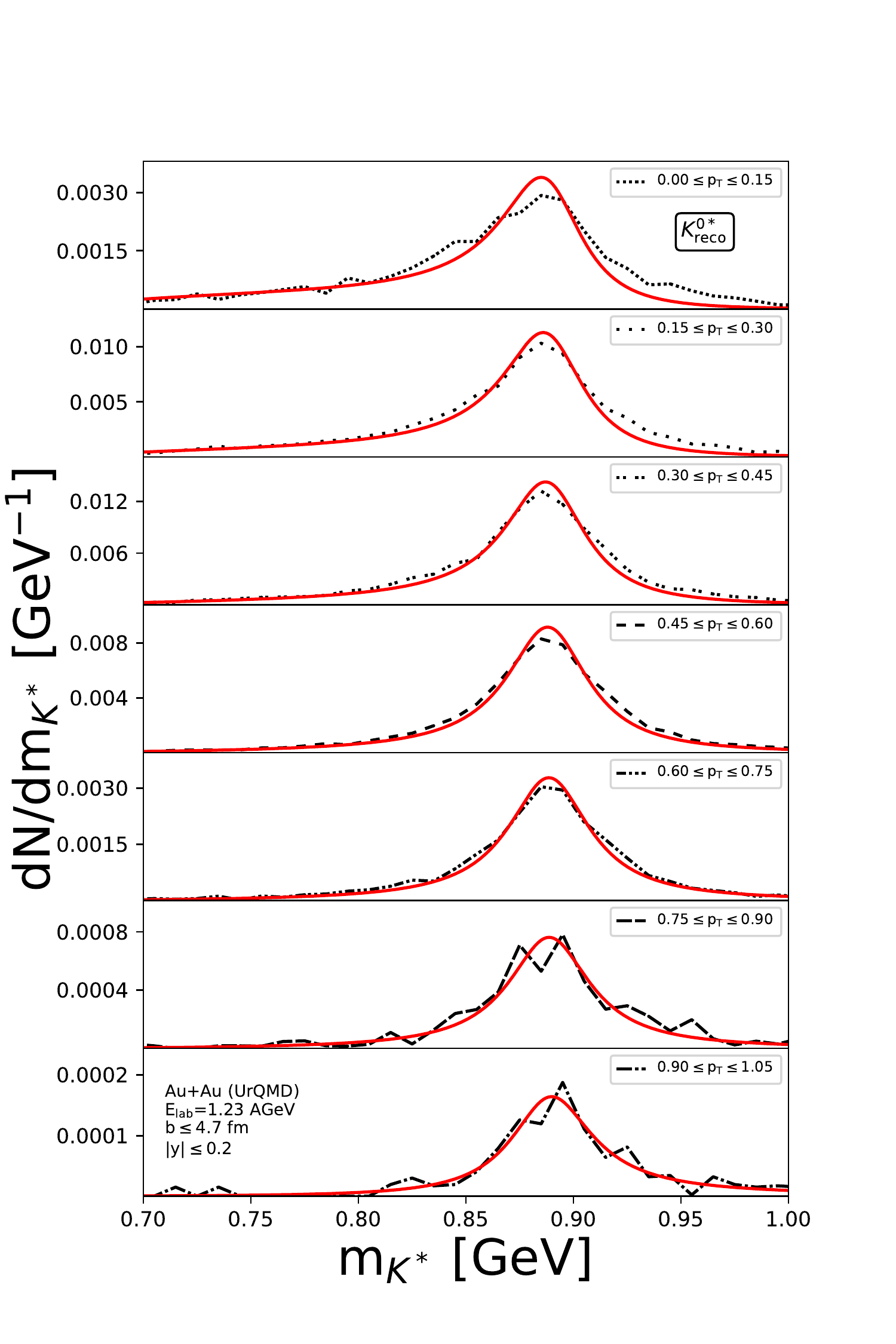}
	\caption{Invariant mass distribution of the reconstructable $K^{0*}$ mesons (black lines) for several transverse momentum bins ranging from 0 to 0.15~GeV at the top to 0.9 to 1.05~GeV at the bottom at mid-rapidity in central Au+Au collisions at $E_\mathrm{beam}=1.23~A$GeV from UrQMD. Also shown are fits to the distributions following the BW$\times$PS argument as red lines.\label{kstar_dndm}}
\end{figure}

To validate the regeneration and freeze-out scenario suggested above, we show in Fig.s \ref{rho_dndm} and \ref{kstar_dndm} the invariant mass distributions\footnote{Note that the invariant mass is calculated from the final momenta of the escaping daughter particles including eventual elastic rescattering reactions, i.e. $m_\mathrm{inv}=\sqrt{(p_1^\mu+p_2^\mu)(p_{1,\mu}+p_{2,\mu})}$.} of the $\rho^0$ and $K^{0*}$ resonances, for several transverse momentum bins (ranging from the lowest $p_\mathrm{T}$ at the top to the highest $p_\mathrm{T}$ at the bottom) as black lines. Also shown are fits to the distributions following the BW$\times$PS argument as red lines. The analysis is done for resonances at mid-rapidity from Au+Au collisions at $E_\mathrm{beam}=1.23~A$GeV from UrQMD.

The invariant mass distribution of the $\rho^0$ meson (Fig. \ref{rho_dndm}, black lines) shows an interesting and pronounced behavior: At high transverse momenta the reconstructed $\rho^0$ masses are populated near the nominal mass $\approx$~770~MeV whereas $\rho^0$s that are reconstructed at low transverse momenta populate the mass region around $\approx$~300~MeV. At intermediate transverse momenta both maxima are overlapping and thus the distribution broadens strongly. There is obviously a kinematic lower boundary for the invariant mass due to the reconstruction in the $\pi^++\pi^-$ channel leading to a trivial narrowing at low $p_{\rm T}$ that may overshadow a broadening due to conventional in-medium effects \cite{vanHees:2004vt}. Bearing in mind the transverse momentum dependence of the reconstruction efficiency, one can clearly observe that the longer a $\rho$ participates in the bulk evolution the later its detectable decay will happen and the lower will be its invariant mass. 

Let us contrast the strongly shifted $\rho^0$ spectral function with the analysis of the $K^{0*}$ resonance mass distribution (Fig. \ref{kstar_dndm}). Here, one does not observe a huge mass shift or substantial broadening of the spectral function. This is in line with our expectations that a $K^*$ once produced and decayed cannot participate in a regeneration cycle and should freeze-out at rather high temperatures. Thus, both resonances confirm our expectations, we will quantify the difference in the emission temperature below.

\subsection{Transverse momentum dependence of the mass shift\label{m_pt}}
\begin{figure} [t!]
	\includegraphics[width=\columnwidth]{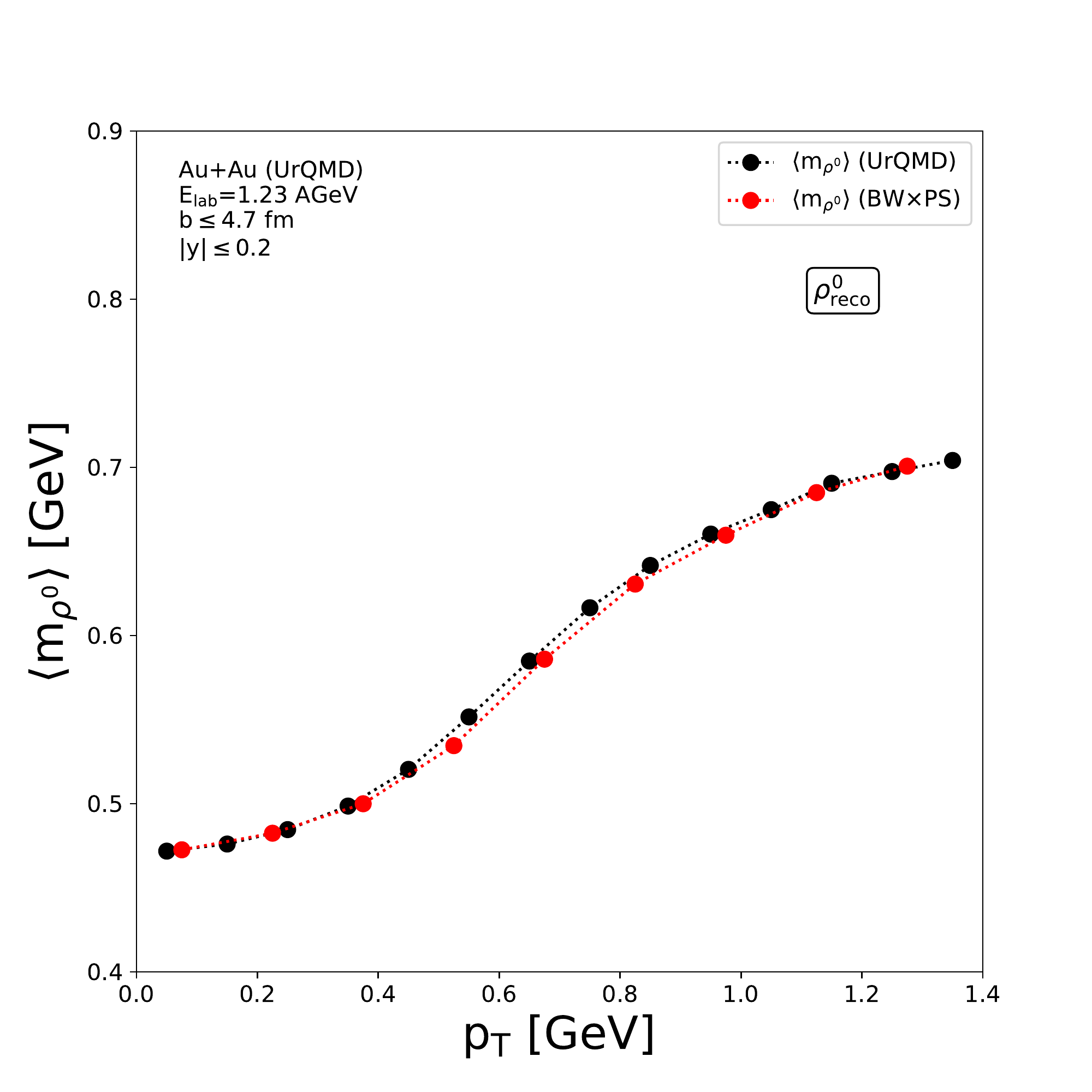}
	\caption{Mean mass of the $\rho^0$ resonance as a function of transverse momentum (black line with full circles) at mid-rapidity in central Au+Au collisions at $E_\mathrm{beam}=1.23~A$GeV from UrQMD and the mean mass calculated by the BW$\times$PS (red line with full squares).\label{rho_meanm_pt}}
\end{figure}

\begin{figure} [t!]
	\includegraphics[width=\columnwidth]{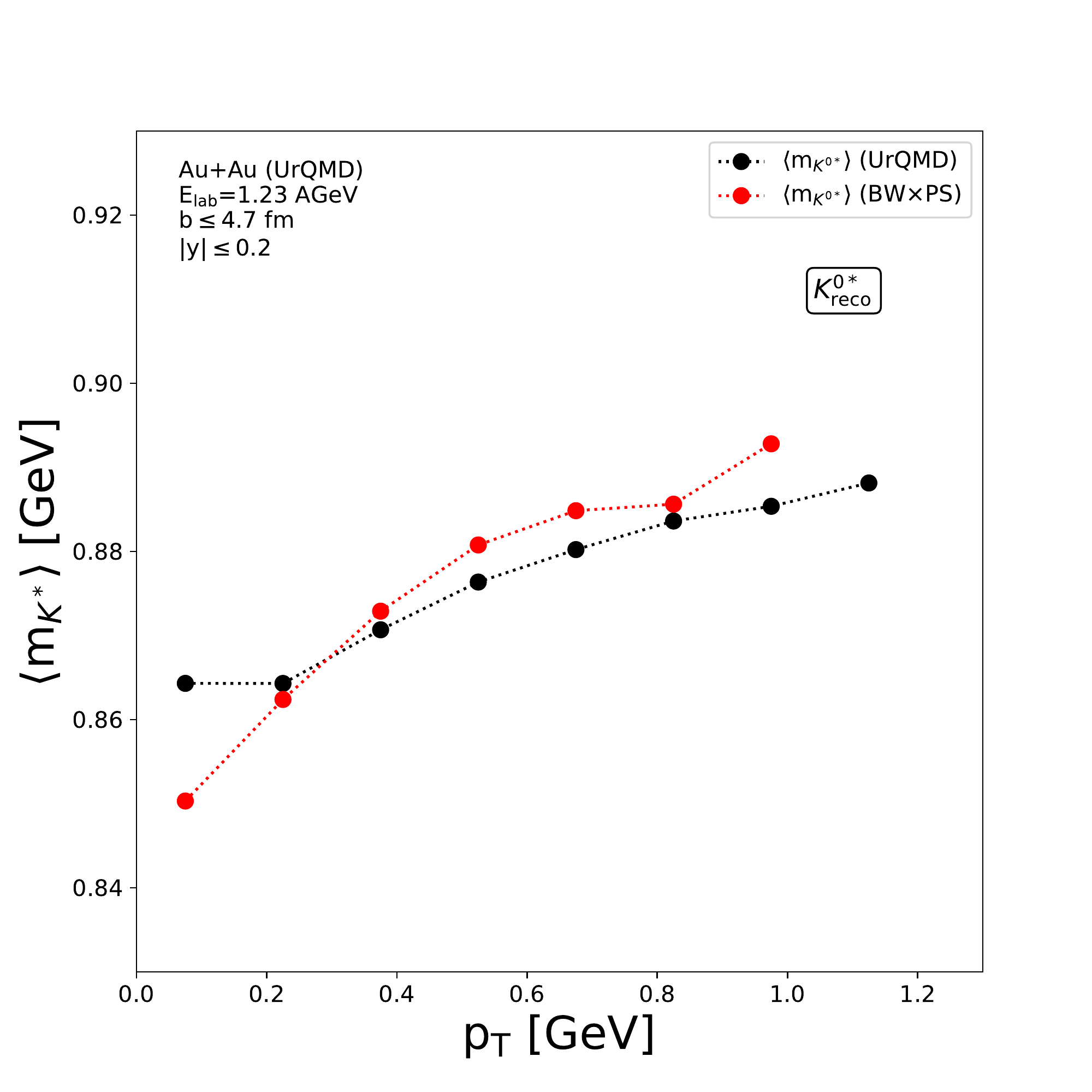}
	\caption{Mean mass of the $K^{0*}$ resonance as a function of transverse momentum (black line with full circles) at mid-rapidity in central Au+Au collisions at $E_\mathrm{beam}=1.23~A$GeV from UrQMD and the mean mass calculated by the BW$\times$PS (red line with full squares).\label{kstar_meanm_pt}}
\end{figure}

Finally, we explore the mass shift of the analyzed resonances in dependence of the transverse momentum shown in Fig. \ref{rho_meanm_pt} for the $\rho^0$ meson and Fig. \ref{kstar_meanm_pt} for the $K^{0*}$ resonance. The black lines show the results from the full transport simulation while the red lines show the average mass from the fitted BW$\times$PS distribution. The average invariant masses obtained in the UrQMD simulation decrease with decreasing transverse momentum. The mean masses obtained by the BW$\times$PS approximation follow the same trend and are obtained by fitting the temperature to the invariant mass distribution in each p$_\mathrm{T}$ bin. For the $\rho$ meson, the BW$\times$PS model captures the trends of the reconstructed average masses as a function of $p_\mathrm{T}$ excellently. This underpins the suggested phase space origin of the mass shift in the case of short-lived resonances. For the $\rho^0$ meson (Fig. \ref{rho_meanm_pt}) we extract a substantial shift of the mean mass by $\approx300$~MeV ($\approx3\Gamma_\rho^\mathrm{tot}$).

For the strange $K^{0*}$ a maximal mass shift of $\approx30$~MeV apart from the nominal mass is predicted. Here, the mean masses obtained from the full UrQMD simulation and from the fitted BW$\times$PS distribution agree very well and differ only by 10~MeV. The absence of a resonant kaon+nucleon interaction clearly suppresses the regeneration capability for $K^*$ resonances thus leaving its invariant mass consistent with the initial temperature. This finding is qualitatively also consistent with flavor hierarchy explored at chemical freeze-out e.g. in \cite{Reichert:2022qvt}.


\subsection{Transverse momentum dependence of the kinetic decoupling temperature}
\begin{figure} [t!]
	\includegraphics[width=\columnwidth]{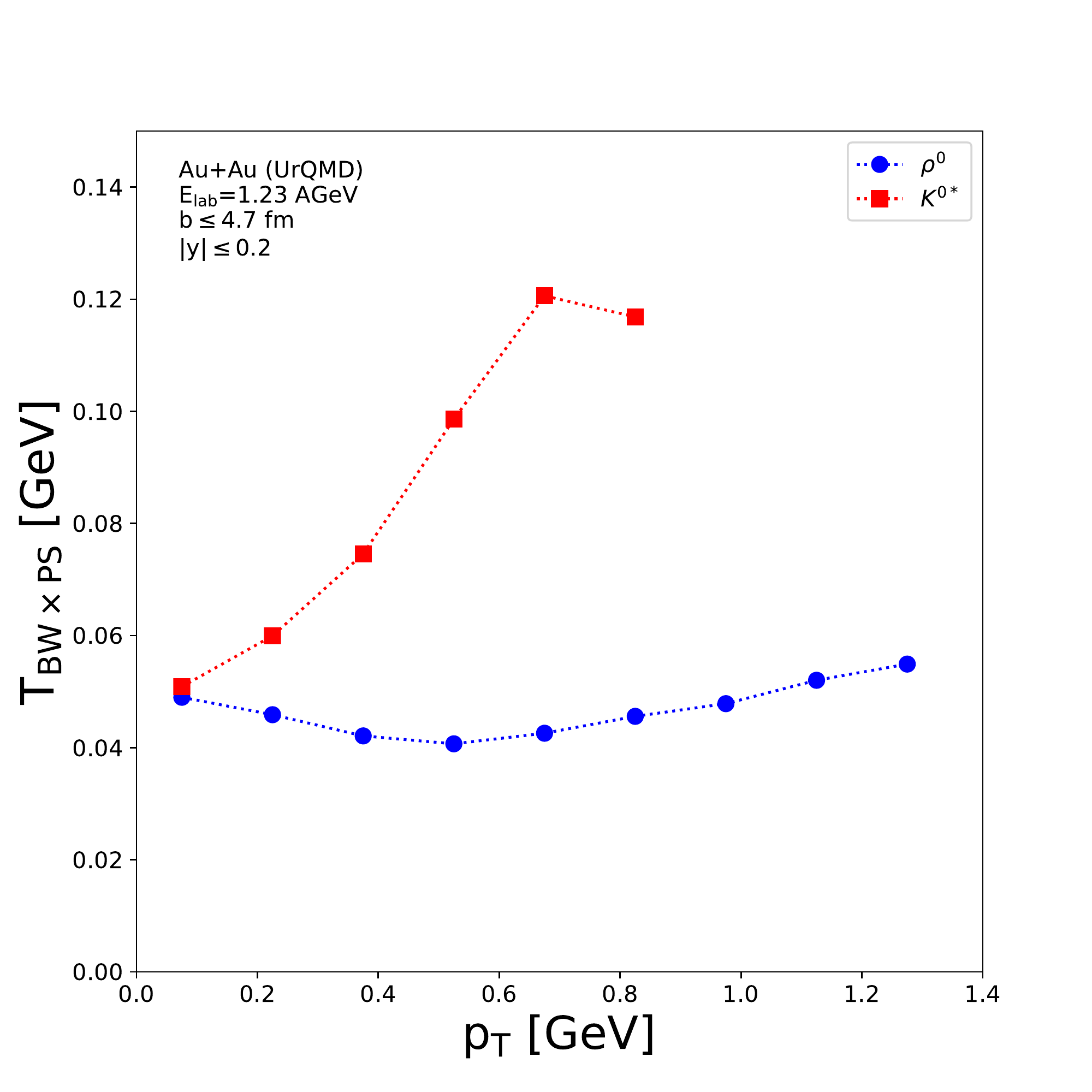}
	\caption{Kinetic decoupling temperature of the $\rho^0$ (blue circles) and the $K^{0*}$ (red squares) resonance as a function of transverse momentum at mid-rapidity in central Au+Au collisions at $E_\mathrm{beam}=1.23~A$GeV from UrQMD.\label{temp_pt_fit}}
\end{figure}

Finally, we can use the mass shift as a thermometer to analyze the temperature at the decoupling stage. The temperatures are obtained by fitting the BW$\times$PS with mass dependent total decay width to the invariant mass distributions in each transverse momentum bin. In Fig. \ref{temp_pt_fit} we show the kinetic decoupling temperature of the $\rho^0$ (blue circles) and the $K^{0*}$ (red squares) resonance as a function of transverse momentum at mid-rapidity in central Au+Au collisions at $E_\mathrm{beam}=1.23~A$GeV from UrQMD. In line with our discussion we observe that the $\rho$ meson decouples very late, i.e. after significant regeneration reactions and when the system has already cooled down. This is reflected in a rather constant decoupling temperature of $T_\mathrm{kin}=40-60$~MeV. The extracted temperature is very close to the kinetic freeze-out temperature ($T=70.3$ MeV) extracted by \cite{Motornenko:2021nds,Harabasz:2022rqg}. The kinetic decoupling temperature of the strange $K^*$ shows a different behavior. At low transverse momenta its decoupling temperature is consistent with the $\rho$ meson around 50 MeV. However, with increasing transverse momentum the estimated kinetic decoupling temperature increases rapidly to $T_\mathrm{kin}=120$~MeV indicating an earlier decoupling from the system. This suggests that $K^*$ at very low p$_\mathrm{T}$ participate to some extent also in some regeneration while high p$_\mathrm{T}$ $K^*$ resonances decouple directly after formation.

The observed mass shifts and the extracted temperatures of the investigated $\rho^0$ and $K^{0*}$ resonances are summarized in Table \ref{max_shift}. Reminiscing our argumentation of the mass shift being a kinematic effect occurring due to a long evolving decay and regeneration cycle of the investigated resonances, the temperatures obtained from the fit to the invariant mass distributions of the BW$\times$PS argument with mass dependent partial decay width in each transverse momentum bin should mirror the temperature on the kinetic freeze-out hyper-surface of the given resonance species at the given p$_\mathrm{T}$. 
\begin{table} [h!]
	\centering
	\begin{tabular}{|c|c|c|}
		\hline
		Resonance & Mass shift & Temperature (BW$\times$PS)\\ \hline 
		\hline
		$\rho^0$& -330~MeV & 40-60~MeV\\
		$K^{0*}$& -30~MeV & 50-120~MeV\\
		$\Delta^{++}$& -50~MeV & 81~MeV \\ 
		\hline 
	\end{tabular}
	\caption{Mass shifts (with respect to the nominal mass) at vanishing transverse momentum  (column 2) and the kinetic decoupling temperature extracted from Eqs. (\ref{BWxPS})-(\ref{PS}) (column 3) for the $\rho^0$ and $K^{0*}$ mesons in central Au+Au collisions at $E_\mathrm{beam}=1.23~A$GeV. We also show the results for the $\Delta^{++}$ taken from \cite{Reichert:2019lny,Reichert:2019zab}.}\label{max_shift}
\end{table}

\section{Summary}
The UrQMD model has been used to study possible observable mass shifts of the $\rho^0$(770) and the $K^{0*}$(892) resonances in their hadronic decay channels in central Au+Au collisions at $E_\mathrm{beam}=1.23~A$GeV. We predicted observable mass shifts for all investigated resonances. The largest mass shift, $\Delta m_\rho \approx -330$ MeV was observed for the $\rho^0$(770) meson. It was explained by the long evolving decay and regeneration cycles in which the finally observable resonances decay in a cold environment near the kinetic freeze-out surface. In case of the $K^*$ the mass shift is moderate, $\Delta m_{K^*} \approx -30$ MeV, because such a regeneration cycle is suppressed, because there is no resonance in the $K+N$ channel.

We related the mass shifts to the decoupling temperature using a Breit-Wigner spectral function folded by the thermal weight of each resonance at the freeze-out temperature of this resonance species.  We found that in general, the invariant mass of a hadronically reconstructable resonance a) increases with increasing transverse momenta and b) decreases with decreasing source temperature. 

Our predictions, can be probed using the HADES experiment at GSI or the low energy data from the RHIC-BES experiments.

\begin{acknowledgements}
The authors want to thank Paula Hillmann for fruitful discussions. This article is part of a project that has received funding from the European Union’s Horizon 2020 research and innovation programme under grant agreement STRONG – 2020 - No 824093
\end{acknowledgements}


\end{document}